\documentclass[aps,preprint,groupedaddress]{revtex4-2}

\usepackage{amsmath}
\usepackage{bm}
\usepackage{mathtools}
\usepackage{multirow}
\usepackage{graphicx}
\usepackage{subcaption}
\usepackage{siunitx}
\usepackage{hyperref}

\begin{document}

\title{Information-Preserving SGS model based on the local inter-scale equilibrium hypothesis}

\author{Takeru Hashimoto}
\affiliation{
    Department of Mechanical and Aerospace Engineering,
    Tokyo University of Science,
    2641 Yamazaki, Noda-shi, Chiba 278-8510, Japan
}
\author{Takahiro Tsukahara}
\affiliation{
    Department of Mechanical and Aerospace Engineering,
    Tokyo University of Science,
    2641 Yamazaki, Noda-shi, Chiba 278-8510, Japan
}
\author{Ryo Araki}
\affiliation{
    Department of Mechanical and Aerospace Engineering,
    Tokyo University of Science,
    2641 Yamazaki, Noda-shi, Chiba 278-8510, Japan
}
\email[]{araki.ryo@rs.tus.ac.jp}

\date{\today}

\begin{abstract}
    Large eddy simulation has been widely used to simulate turbulence at balanced computational cost and accuracy.
    Many Subgrid-Scale (SGS) models have been proposed over the years, where data-driven and machine learning-aided approaches set the recent trend.
    To address the problem of extrapolation in these models, we propose a new data-driven SGS model based on an information-theoretic picture of turbulence.
    To this end, we estimate the model parameters by maximizing mutual information, which correspond to the scale-by-scale local equilibrium hypothesis in developed turbulence or ``information preservation.''
    An \textit{a priori} test confirmed that the estimated parameters are in good agreement with the previously reported empirical values.
    Furthermore, \textit{a posteriori} tests on periodic box turbulence and channel turbulence exhibited accuracy comparable to the existing models.
    These results suggest the utility of the information-theoretic picture of turbulence for constructing more generic SGS models without the need for empirically prescribed model parameters, while enhancing physical interpretability beyond black-box approaches.
\end{abstract}

\maketitle

\section{Introduction}
\label{sec:Introduction}

Turbulence has a significant influence on our daily lives, and its numerical simulation plays a crucial role in various engineering fields.
However, Direct Numerical Simulation (DNS), which solves vortices over a wide range of scales, provides a precise representation of turbulence but is too computational demanding for industrial flow simulations.
Instead, we conduct low-cost simulation of turbulence by Reynolds-Averaged Navier--Stokes simulation (RANS) or Large Eddy Simulation (LES), where the computational cost is significantly reduced~\cite{Pope:2001, Davidson:2015}.
RANS only solves the time-averaged governing equations to reproduce the steady component of the flow, but it cannot capture its unsteady nature.
In contrast, LES reduces the computational cost by solving the spatially filtered governing equations and can reproduce the unsteady flow field.

LES requires the Subgrid-Scale (SGS) model to represent the motion of the filtered small scales.
Many SGS models, including the Smagorinsky model~\cite{Smagorinsky:1963}, have been proposed and reviewed in many publications~\cite{Lesieur:1996,Meneveau:2000, Yang:2015,Kajishima:2016, Moser:2021}.
In recent years, data-driven SGS models employing machine learning framework have attracted significant attention~\cite{Gamahara:2017, Brunton:2020, Taira:2025,Beck:2021, Duraisamy:2021,Sharma:2023, Zhang:2023}.
One of the key difficulty in these models is related with the extrapolation or the generalizability to flows different from the training data~\cite{Park:2021, Yang:2024}.
For example, Inubushi \textit{et al.}~\cite{Inubushi:2020} discussed the validity of the transfer learning, where a model trained on low Reynolds number flows can be applied to higher Reynolds numbers, owing to the universality associated with the energy cascade of turbulence.
Subel \textit{et al.}~\cite{Subel:2021} showed that a model trained by low Reynolds number flows can achieve stable predictions with a few of high-Reynolds-number data by retraining only the layers near the output of the deep learning model.
However, when it comes to more intense turbulence (e.g, associated with airplanes), it is extremely difficult to obtain the ``ground truth'' data to be used in training from neither simulation nor experiment.
Moreover, such flows often contain extreme events which may hinder data-driven modeling~\cite{Brunton:2020, Fukami:2023}.

Recently, apart from the machine learning context, information theory has been employed in research to characterize turbulence.
Boffetta \textit{et al.}~\cite{Boffetta:2002} and subsequent studies attempted to characterize the chaotic nature of turbulence by information-theoretic quantities.
One example is the statistical irreversibility of energy cascade characterized by information entropy~\cite{Vela-Martin:2021}.
Lozano-Durán and Arranz~\cite{Lozano:2022} defined a quantity called information flux as an extension of transfer entropy, to investigate the causal nature of turbulence associated with energy cascade, which later refined by Lagrangian perspective~\cite{Araki:2024}.
For the laminar--turbulent transition, information-theoretic quantities can identify different stages~\cite{Bahamonde:2023}.
Alternatively, information thermodynamics has been employed to connect the microscopic thermal fluctuations of the fluid molecules and turbulence statistics.
Tanogami and Araki~\cite{Tanogami:2024, Tanogami:2025} quantified the flow of information in three-dimensional turbulence from large to small scales from the second law of information thermodynamics.

There are several studies where information theory is applied on LES.
Engelmann~\cite{Engelmann:2022} found the Shannon entropy to increases in low-resolution simulations, and claimed that this quantity is a good indicator of sufficient resolution in LES.
Vela-Martín~\cite{Vela-Martín:2025} found similarities between the LES formulation and the so-called ``uncertainty cascade.''
Lozano-Durán and Arranz~\cite{Lozano:2022} proposed a SGS model which estimates the model parameters by Kullback--Leibler (KL) divergence.
More specifically, they evaluated the similarity between the PDF (probability density function) of the energy flux at two different scales by KL divergence to determine the model parameters.
Their Information-Preserving SGS (IP-SGS) model serves as the starting point of this study.

\begin{figure}[tb]
    \centering
    \includegraphics[width=0.8\textwidth]{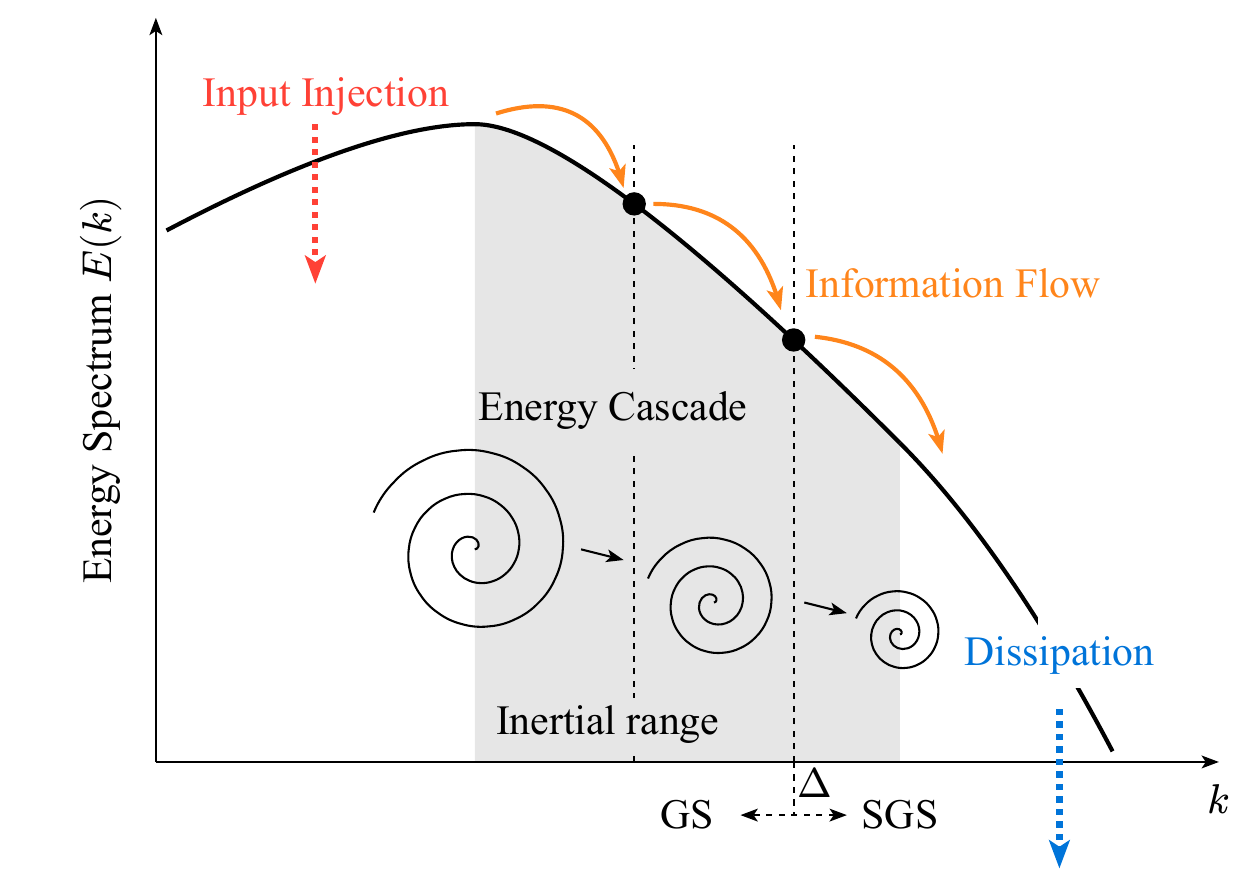}
    \caption{
        Schematic of the energy spectrum in developed turbulence.
        Energy injected at large scales transfers to smaller scales via scale-local nonlinear interactions, before dissipated at the viscous scales.
        Recently, information of large-scale quantities is also found to propagate towards smaller scales~\cite{Lozano:2022,Tanogami:2024}.
        LES resolves the grid scale (GS) $\Delta$ within the inertial range by modeling the non-resolved SGS $\ell < \Delta$.
    }
    \label{fig:schematic_energy_spectrum}
\end{figure}

In this study, we propose a SGS model based on the ``information preservation'' concept by combining recent information-theoretic approaches with the local inter-scale equilibrium hypothesis.
Previous studies~\cite{Lozano:2022,Tanogami:2024} indicated that information associated with the large-scale eddies in three-dimensional turbulence propagates from large to small scales with scale-local manner, similar to the energy cascade (see Fig.~\ref{fig:schematic_energy_spectrum}).
In addition, turbulence does not transfer new information from small to large scales but only redundant information that exists in both scales~\cite{Martinez-Sanchez:2024}.
These results suggest that an SGS model should reproduce the information transfer from the grid scale (GS) to the SGS.
To explicitly incorporate this feature, we propose a new SGS model that maximizes the correlation between the GS and a larger scale, thereby achieving information preservation between the two scales.
Note that we assume the two scales are not far apart and ignore the time delay involved in information transfer.
Our model is inspired by the recently proposed IP-SGS model~\cite{Lozano:2022}.
The key improvement is that while the IP-SGS relies on scaling laws obtained from DNS, our model is based on the local inter-scale equilibrium hypothesis to avoid reliance on known scaling laws.

This paper is organized as follows.
Section~\ref{sec:Information-Preserving_Coherent_Structure_Model} introduces an Information-Preserving Coherent Structure Model (IP-CSM).
Section~\ref{sec:A_Posteriori_test_of_the_IP-CSM} reports \textit{a posteriori} tests of the proposed model for periodic box turbulence and channel turbulence.
Section~\ref{sec:Discussion_and_Conclusion} discusses the characteristics of the proposed model and incorporates it in the context of data-driven modeling approaches in turbulence.

\section{Information-Preserving Coherent Structure Model}
\label{sec:Information-Preserving_Coherent_Structure_Model}

In this section, we propose an IP-CSM.
Section~\ref{subsec:LES_formulation} describes the general formulation of SGS modeling, before introducing the fundamentals of information theory in Sec.~\ref{subsec:Information_theoretic_quantities}.
Section~\ref{subsec:Model_term_selection_via_a_priori_test}--\ref{subsec:Model_parameters_estimation_based_on_local_scale-by-scale_equilibrium_hypothesis} present the formulation of the proposed IP-CSM via three steps: model term selections in Sec.~\ref{subsec:Model_term_selection_via_a_priori_test}, invariant model parameter definitions in Sec.~\ref{subsec:Invariant_parameter_determination_using_coherent_structure_function}, and temporal parameter estimations in Sec.~\ref{subsec:Model_parameters_estimation_based_on_local_scale-by-scale_equilibrium_hypothesis}.

\subsection{LES formulation}
\label{subsec:LES_formulation}

We consider incompressible fluid governed by the continuity equation
\begin{equation}
    \frac{\partial u_i}{\partial x_i} = 0
    \label{eq:continuity_equation}
\end{equation}
and the Navier--Stokes equations
\begin{equation}
    \frac{\partial u_i}{\partial t} + u_j\frac{\partial u_i}{\partial x_j} = - \frac{1}{\rho} \frac{\partial \Pi}{\partial x_i} + \nu \frac{\partial^2 u_i}{\partial x_j \partial x_j} + f_i.
    \label{eq:momentum_equation}
\end{equation}
Here, $u_i$, $\Pi$, and $f_i$ denote the velocity, pressure, and external force, respectively.
The flow is parametrized by the density $\rho$ and the kinematic viscosity $\nu$.
In the LES formulation, we numerically solve a spatially coarse-grained velocity field
\begin{equation}
    \bar{u}_i(\boldsymbol{x}) \coloneq \int_V G(\boldsymbol{x} - \boldsymbol{x}'; \Delta) u_i(\boldsymbol{x}') \,\mathrm{d}\boldsymbol{x}',
\end{equation}
where $G(\boldsymbol{x}; \Delta)$ is a filter function with the filter width $\Delta$.
Note that $\overline{(\cdot)}$ denotes the filtering.
The filter function must have the following characteristics: positivity near the origin $G(\boldsymbol{x}; \Delta) > 0$ near $\boldsymbol{x} = \boldsymbol{0}$, boundedness $\lim_{|\boldsymbol{x}|\to\infty} G(\boldsymbol{x}; \Delta) = 0$, and regularity $\int_V G(\boldsymbol{x}; \Delta) \,\mathrm{d}\boldsymbol{x} = 1$.
Among with typical filter functions, such as top-hat and spectral-cutoff, we employ the Gaussian filter
\begin{align}
    G(\boldsymbol{x};\Delta) & \coloneq \prod_{i=1}^3 G_i(x;\Delta),                                           \\
    G_i(x;\Delta)            & \coloneq \sqrt{\frac{6}{\pi \Delta^2}} \exp\left(-\frac{6x^2}{\Delta^2}\right),
\end{align}
throughout this study.
Note that similar results hold for the other filter functions.
After coarse-graining Eqs.~\eqref{eq:continuity_equation} and \eqref{eq:momentum_equation}, we obtain the filtered continuity equation
\begin{equation}
    \frac{\partial \bar{u}_i}{\partial x_i} = 0
    \label{eq:filtered_continuity_equation}
\end{equation}
and the filtered Navier--Stokes equations
\begin{equation}
    \frac{\partial \bar{u}_i}{\partial t} + \bar{u}_j\frac{\partial \bar{u}_i}{\partial x_j}  = - \frac{1}{\rho} \frac{\partial \bar{\Pi}}{\partial x_i} + \nu \frac{\partial^2 \bar{u}_i}{\partial x_j \partial x_j} - \frac{\partial \tau_{ij}}{\partial x_j} + \bar{f}_i
    \label{eq:filtered_momentum_equation}
\end{equation}
associated with the SGS stress tensor
\begin{equation}
    \tau_{ij} \coloneq \overline{u_i u_j} - \bar{u}_i \bar{u}_j
    \label{eq:sgs_stress_tensor}
\end{equation}
resulted from the filtered nonlinear term.
In LES, we numerically solve Eqs.~\eqref{eq:filtered_continuity_equation}, \eqref{eq:filtered_momentum_equation}, and modeled \eqref{eq:sgs_stress_tensor}, to simulate turbulent flows with much smaller computational cost in compare to DNS which directly solves Eqs.~\eqref{eq:continuity_equation} and \eqref{eq:momentum_equation}.

The SGS stress requires modeling as we cannot directly compute $\tau_{ij}$ from the filtered or GS quantities.
In this study, we express the SGS stress  with the following five terms~\cite{Lund:1993}
\begin{align}
    \tau_{ij}^{\mathrm{mod}} & =
    \underbrace{C_1 f_1 \Delta^2 |\boldsymbol{S}| \bar{S}_{ij}}_{M_{ij}^{[1]}}
    + \underbrace{C_2 f_2 \Delta^2 \bar{S}_{ik} \bar{S}_{kj}}_{M_{ij}^{[2]}}
    + \underbrace{C_3 f_3 \Delta^2 \bar{\Omega}_{ik} \bar{\Omega}_{kj}}_{M_{ij}^{[3]}}\nonumber                                                       \\
                             & \quad + \underbrace{C_4 f_4 \Delta^2 (\bar{S}_{ik} \bar{\Omega}_{kj} - \bar{\Omega}_{ik} \bar{S}_{kj})}_{M_{ij}^{[4]}}
    + \underbrace{C_5 f_5 \frac{\Delta^2}{|\boldsymbol{S}|}(\bar{S}_{ik}\bar{S}_{kl}\bar{\Omega}_{lj} - \bar{\Omega}_{ik}\bar{S}_{kl}\bar{S}_{lj})}_{M_{ij}^{[5]}},
    \label{eq:sgs_stress_model}
\end{align}
based on the filtered rate-of-strain tensor
\begin{equation}
    \bar{S}_{ij} \coloneq \frac{1}{2} \left( \frac{\partial \bar{u}_i}{\partial x_j} + \frac{\partial \bar{u}_j}{\partial x_i} \right)
\end{equation}
and the rate-of rotation tensor
\begin{equation}
    \bar{\Omega}_{ij} \coloneq \frac{1}{2} \left( \frac{\partial \bar{u}_i}{\partial x_j} - \frac{\partial \bar{u}_j}{\partial x_i} \right),
\end{equation}
with $|\boldsymbol{S}| \coloneq \sqrt{2 \bar{S}_{ij} \bar{S}_{ij}}$.

The physical interpretation of each term in Eq.~\eqref{eq:sgs_stress_model} is given as follows.
The first term $M_{ij}^{[1]}$ shows a strong similarity with the energy production $\tau_{ij} \bar{S}_{ij}$ while a weak similarity with the SGS stress $\tau_{ij}$~\cite{Meneveau:2000,Horiuti:1997}.
In contrast, $M_{ij}^{[2]}$, $M_{ij}^{[3]}$, and $M_{ij}^{[4]}$ exhibit a stronger similarity with $\tau_{ij}$ compared to $M_{ij}^{[1]}$ as they constitute the Clark term.
Among them, $M_{ij}^{[2]}$ and $M_{ij}^{[3]}$ may induce backscatter of energy and potentially destabilize the SGS model, whereas $M_{ij}^{[4]}$ and $M_{ij}^{[5]}$ do not produce energy and only modify the correlation with $\tau_{ij}$.

The present model involves two kinds of model parameters: $f_i(\boldsymbol{x}, t)$ and $C_i(t)$.
The spatio-temporally varying $f_i(\boldsymbol{x},t)$ is defined by the invariants of $S_{ij}$ and $\Omega_{ij}$~\cite{Lund:1993}:
\begin{align}
    I_1 & = \bar{S}_{ij} \bar{S}_{ji},
        & I_2                                                 & = \bar{\Omega}_{ij} \bar{\Omega}_{ji},
        & I_3                                                 & = \bar{S}_{ij} \bar{S}_{jk} \bar{S}_{ki},\nonumber                                                   \\
    I_4 & = \bar{S}_{ij} \bar{\Omega}_{jk} \bar{\Omega}_{ki},
        & I_5                                                 & = \bar{S}_{ij} \bar{S}_{jk} \bar{\Omega}_{kl} \bar{\Omega}_{li},
        & I_6                                                 & = \bar{\Omega}_{ij} \bar{S}_{jk} \bar{\Omega}_{kl} \bar{\Omega}_{lm} \bar{S}_{mn} \bar{\Omega}_{ni},
\end{align}
and determined in Sec.~\ref{subsec:Invariant_parameter_determination_using_coherent_structure_function}.
The time-dependent $C_i(t)$ is determined by information theory in Sec.~\ref{subsec:Model_parameters_estimation_based_on_local_scale-by-scale_equilibrium_hypothesis}.

\subsection{Information theoretic quantities}
\label{subsec:Information_theoretic_quantities}

Originally proposed by Shannon~\cite{Shannon:1948}, information theory is a theory of communication which provides a mathematical framework of ``information'', a concept often used ambiguously in our daily life.
The key quantity is the Shannon entropy
\begin{equation}
    H(X) \coloneq - \int_{-\infty}^{\infty} p(x) \log p(x) \,\mathrm{d}x,
    \label{eq:shannon_entropy}
\end{equation}
which measures the uncertainty of a continuous random variable $X$.
Here, $p(x)$ denotes the probability density function of $X$.
The joint Shannon entropy
\begin{equation}
    H(X,Y) \coloneq - \int_{-\infty}^{\infty} \int_{-\infty}^{\infty} p(x,y) \log p(x,y) \,\mathrm{d}x \,\mathrm{d}y
    \label{eq:joint_entropy}
\end{equation}
serves as an extension of $H(X)$ to two random variables $X$ and $Y$.

To quantify the amount of shared information between two random variables, we define the mutual information
\begin{equation}
    I(X:Y) \coloneq H(X) + H(Y) - H(X,Y)
    \label{eq:mutual_information}
\end{equation}
representing the statistical dependence between two random variables.
Unlike the correlation coefficient, the mutual information is capable of capturing nonlinear correlation~\cite{Tohme:2024}.
Note that $I(X:Y) = 0$ is attained only when $X$ and $Y$ are statistically independent, and larger values indicate stronger dependence between them.
Furthermore, the mutual information and the Kullback--Leibler (KL) divergence~\cite{Kullback:1951}, a quantity widely used in the context of machine learning, are related by
\begin{equation}
    \mathrm{KL}(p(x,y) \| p(x)p(y)) \coloneq \int_{-\infty}^{\infty} \int_{-\infty}^{\infty} p(x,y) \log \left(\frac{p(x,y)}{p(x)p(y)}\right) \,\mathrm{d}x \,\mathrm{d}y = I(X:Y).
    \label{eq:kl_divergence}
\end{equation}
This relation infers that the maximization of mutual information can be interpreted as maximizing the KL divergence between the joint probability distribution $p(x,y)$ and the product of the marginal distributions $p(x)p(y)$.

\subsection{Model term selection via a priori test}
\label{subsec:Model_term_selection_via_a_priori_test}

During the SGS model construction, we select one or two terms from Eq.~\eqref{eq:sgs_stress_model}, often heuristically, to keep the model simple.
For example, both the original Smagorinsky model~\cite{Smagorinsky:1963} and the coherent structure model~\cite{Kobayashi:2005} employ only the first term $M_{ij}^{[1]}$.
Some other models, such as the vortex-stretching-based nonlinear model~\cite{Silvis:2020} and the information-preserving SGS model~\cite{Lozano:2022} select two terms: $M_{ij}^{[1]}$ and $M_{ij}^{[4]}$.

The above choices were supported by Lund and Novikov~\cite{Lund:1993} using a correlation-based analysis between DNS data and modeled SGS stress.
They estimated the parameters $f_i(\boldsymbol{x}, t)$ and $C_i(t)$ of each model term $M_{ij}^{[n]}$ by the least-squares method based on the ``true'' SGS stress $\tau_{ij}$ (Eq.~\eqref{eq:sgs_stress_tensor}) computed from DNS.
They then evaluated the correlation coefficient
\begin{equation}
    \mathrm{Cor}(\tau_{ij}, \tau_{ij}^\mathrm{mod}) \coloneq \frac{\langle\tau_{ij}\tau_{ji}^{\mathrm{mod}}\rangle}{\sqrt{\langle\tau_{ij}\tau_{ij}\rangle\langle\tau_{ji}^{\mathrm{mod}}\tau_{ji}^{\mathrm{mod}}\rangle}}
\end{equation}
between the true SGS stress $\tau_{ij}$ and the modeled SGS stress $\tau_{ij}^{\mathrm{mod}}$, which consists of a linear combination of one to five model terms $M_{ij}^{[n]}$.
They obtained the highest correlation with $\tau_{ij}^\mathrm{mod} = M_{ij}^{[1]}$ for the one-term model, and $\tau_{ij}^\mathrm{mod} = M_{ij}^{[1]} + M_{ij}^{[4]}$ for the two-term model.

In this study, we re-evaluate the above result from information-theoretic point of view.
To this end, we focus not on the the SGS stress $\tau_{ij}$ itself as in~\cite{Lund:1993}, but on the force $\partial_j\tau_{ij}$ appeared in the filtered Navier--Stokes equations~\eqref{eq:filtered_momentum_equation} and the energy flux (or energy production in the SGS) $\tau_{ij}\bar{S}_{ij}$ appeared in the energy equation (to be defined at~\eqref{eq:tg_energy_equation}).
To keep our model simple, we select two terms from Eq.~\eqref{eq:sgs_stress_model} to model the SGS stress as
\begin{equation}
    \tau_{ij}^{\mathrm{mod}} =M_{ij}^{[m]} + M_{ij}^{[n]}, \quad m, n \in \{1, 2, 3, 4, 5\}.
    \label{eq:sgs_model_selected}
\end{equation}
Then, we calculate the mutual information
\begin{equation}
    I\left(\left[\frac{\partial \tau_{ij}}{\partial x_j}, \tau_{ij}\bar{S}_{ij}\right]: \left[\frac{\partial \tau_{ij}^{\mathrm{mod}}}{\partial x_j}, \tau_{ij}^{\mathrm{mod}}\bar{S}_{ij}\right]\right)
    \label{eq:mi_sgs_force_energy}
\end{equation}
between the DNS-based $\tau_{ij}$ (see Eq.~\eqref{eq:sgs_stress_tensor}) and the model-based $\tau_{ij}^{\mathrm{mod}}$ (see Eq.~\eqref{eq:sgs_model_selected}).
We note that the highest $I$ infers the most ``informative'' model.
At this point, we followed \citet{Lund:1993} to determine the model parameters $f_i(\boldsymbol{x}, t) C_i(t)$ (without the Einstein rule) by the least-squares method, without distinguishing $f_i$ and $C_i$.
More specifically, the above model coefficients are obtained by minimizing $\left[\tau_{ij}(\boldsymbol{x}, t) - \tau_{ij}^{\mathrm{mod}}(\boldsymbol{x}, t)\right]^2$.

\begin{table}[tbp]
    \centering
    \caption{
        Parameters and statistical quantities of the periodic box turbulence.
        The grid resolution $N^3$;
        the Taylor-scale-based Reynolds number $\mathrm{Re}_\lambda$;
        the number of samples to compute the statistics $N_\mathrm{sample}$;
        the time step width $\Delta t$;
        and the total simulation time $T_\mathrm{sim}$.
        Note that both $\Delta t$ and $T_\mathrm{sim}$ are normalized by the large-eddy turnover time $T_\mathrm{turn} = L_e / U_e$.
        Here, $L_e$ and $U_e$ are the mean energy length and the mean energy velocity, respectively.
        The large-eddy turnover time $T_\mathrm{turn}$ is defined using DNS and is also used for the nondimensionalization of the LES.
    }
    \begin{tabular}{cccccc} \hline\hline
            & $N^3$   & $\mathrm{Re}_\lambda$ & $N_\mathrm{sample}$ & $\Delta t /T_\mathrm{turn}$          & $T_\mathrm{sim} / T_\mathrm{turn}$ \\ \hline
        DNS & $256^3$ & \multirow{2}{*}{170}  & \multirow{2}{*}{15} & \multirow{2}{*}{$5.8\times 10^{-4}$} & \multirow{2}{*}{5.8}               \\
        LES & $64^3$  &                       &                     &                                      &                                    \\ \hline
    \end{tabular}
    \label{tab:periodic_box_turbulence}
\end{table}

For the model term selection, we conducted DNS of a periodic box turbulence in $(2\pi)^3$ box, a canonical configuration of developed turbulence away from walls.
We utilized an open-source solver Incompact3d~\cite{Laizet:2009, Paul:2020}.
To attain a statistically steady turbulence, we forced the flow by the four-vortices Taylor--Green force
\begin{equation}
    (f_1, f_2, f_3) = (- \sin x_1 \cos x_2, \cos x_1 \sin x_2, 0).
    \label{eq:Taylor--Green_force}
\end{equation}
The parameters and some statistical quantities of the DNS are summarized in Table~\ref{tab:periodic_box_turbulence}.

\begin{figure}[tbp]
    \centering
    \includegraphics[width=0.6\textwidth]{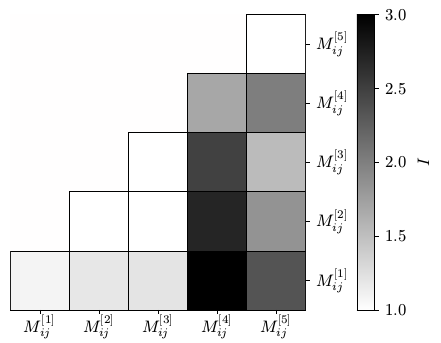}
    \caption{
        Mutual information, Eq.~\eqref{eq:mi_sgs_force_energy}, between the SGS stress computed directly from DNS and the modeled SGS stress consist of two terms.
        The diagonal elements represent models with one term.
    }
    \label{fig:mi_mix}
\end{figure}

Figure~\ref{fig:mi_mix} shows the mutual information (Eq.~\eqref{eq:mi_sgs_force_energy}) for all combinations of two terms from Eq.~\eqref{eq:sgs_stress_model}.
Among all combinations,
\begin{equation}
    \tau_{ij}^{\mathrm{mod}}
    = \underbrace{C_1 f_1 \Delta^2 |\boldsymbol{S}| \bar{S}_{ij}}_{M_{ij}^{[1]}}
    + \underbrace{C_4 f_4 \Delta^2 (\bar{S}_{ik} \bar{\Omega}_{kj} - \bar{\Omega}_{ik} \bar{S}_{kj})}_{M_{ij}^{[4]}},
    \label{eq:selected_model}
\end{equation}
yields the largest mutual information.
This result agrees with the correlation-based analysis~\cite{Lund:1993} and the conventional choice employed in the previous studies~\cite{Lozano:2022,Silvis:2020}.
As $M_{ij}^{[1]}$ correlates with the energy production and $M_{ij}^{[4]}$ with the SGS stress, our model (Eq.~\eqref{eq:selected_model}) accurately reproduces both the force and the energy production.
We further discuss the case of three-term models in Appendix~\ref{sec:Appendix_A}.

\subsection{Invariant parameter determination using coherent structure function}
\label{subsec:Invariant_parameter_determination_using_coherent_structure_function}

In this subsection, we define the unknown space-time dependent parameter as
\begin{align}
    f_i(\boldsymbol{x}, t) & \coloneq |F_{\mathrm{CS}}|^{p_i},\label{eq:fi_definition}
\end{align}
using the coherent structure function~\cite{Kobayashi:2005}
\begin{align}
    F_{\mathrm{CS}} & \coloneq \frac{\bar{S}_{ij}\bar{S}_{ij} - \bar{\Omega}_{ij}\bar{\Omega}_{ij}}{\bar{S}_{kl}\bar{S}_{kl} + \bar{\Omega}_{kl}\bar{\Omega}_{kl}}.
    \label{eq:coherent_structure_function}
\end{align}
It is a dimensionless representation of the second invariant of the velocity gradient tensor (commonly referred to as the $Q$ value), thus bounded within the range of $-1 \leq F_{\mathrm{CS}} \leq 1$.
By using $F_{\mathrm{CS}}$, one can reproduce two flow behaviors: $\tau_{ij} = 0$ in laminar and near-wall regions.

We determine the exponent $p_i$ in Eq.~\eqref{eq:fi_definition} from the near-wall asymptotic condition.
In the vicinity of a solid wall, we approximate with a Taylor series expansion of the wall-normal coordinate $y$ as
\begin{align}
    u_1 & = a_1 y + b_1 y^2 + c_1 y^3 + \mathcal{O}(y^4), \label{eq:wall_profile_u1} \\
    u_2 & = b_2 y^2 + c_2 y^3 + \mathcal{O}(y^4), \label{eq:wall_profile_u2}         \\
    u_3 & = a_3 y + b_3 y^2 + c_3 y^3 + \mathcal{O}(y^4). \label{eq:wall_profile_u3}
\end{align}
Note that the zeroth-order terms $\mathcal{O}(y^0)$ vanish due to the Dirichlet boundary condition $u_i = 0$ at the wall, and the $\mathcal{O}(y^1)$ term in Eq.~\eqref{eq:wall_profile_u2} vanishes according to the continuity constraint $\partial_y u_2 = 0$ at the wall.
From these expressions, the coherent structure function and the SGS stress tensor satisfy the near-wall scaling
\begin{equation}
    F_{\mathrm{CS}} = \mathcal{O}(y^2)
    \label{eq:fcs_wall_asymptotic}
\end{equation}
and
\begin{equation}
    \tau_{ij} =
    \begin{cases}
        \mathcal{O}(y^4) \quad (i, j) = (2, 2),                         \\
        \mathcal{O}(y^3) \quad (i, j) = (1, 2), (2, 1), (2, 3), (3, 2), \\
        \mathcal{O}(y^2) \quad \text{otherwise},
    \end{cases}
\end{equation}
respectively.
For the model terms $M_{ij}^{[1]}$ and $M_{ij}^{[4]}$, their leading order is
\begin{align}
    \bar{S}_{ij}                                                  & =
    \begin{cases}
        \mathcal{O}(y^1) \quad (i, j) = (2, 2),                         \\
        \mathcal{O}(y^0) \quad (i, j) = (1, 2), (2, 1), (2, 3), (3, 2), \\
        0 \quad\quad\quad \text{otherwise},
    \end{cases}   \\
    \bar{S}_{ik}\bar{\Omega}_{kj} - \bar{\Omega}_{ik}\bar{S}_{kj} & =
    \begin{cases}
        \mathcal{O}(y^1) \quad (i, j) = (1, 2), (2, 1), (2, 3), (3, 2), \\
        \mathcal{O}(y^0) \quad \text{otherwise}.
    \end{cases}
\end{align}
Thus, the invariant functions must scale as
\begin{equation}
    f_1 (\boldsymbol{x}, t) = \mathcal{O}(y^3)\sim \left|F_{\mathrm{CS}}\right|^{\frac{3}{2}} \quad \text{and} \quad
    f_4 (\boldsymbol{x}, t) \geq \mathcal{O}(y^4)\sim \left|F_{\mathrm{CS}}\right|^{2}.
    \label{eq:fi_wall_asymptotic}
\end{equation}
Note that the inequality for $f_4$ enforces a higher-order scaling which does not violate the asymptotic requirement already satisfied by $f_1$.
To satisfy these inequalities, we set the exponent $p_4 = 2$ in Eq.~\eqref{eq:fi_definition}, which yields $\mathcal{O}(y^4)$.

\subsection{Model parameters estimation based on local scale-by-scale equilibrium hypothesis}
\label{subsec:Model_parameters_estimation_based_on_local_scale-by-scale_equilibrium_hypothesis}

\begin{figure}
    \centering
    \includegraphics[width=0.8\textwidth]{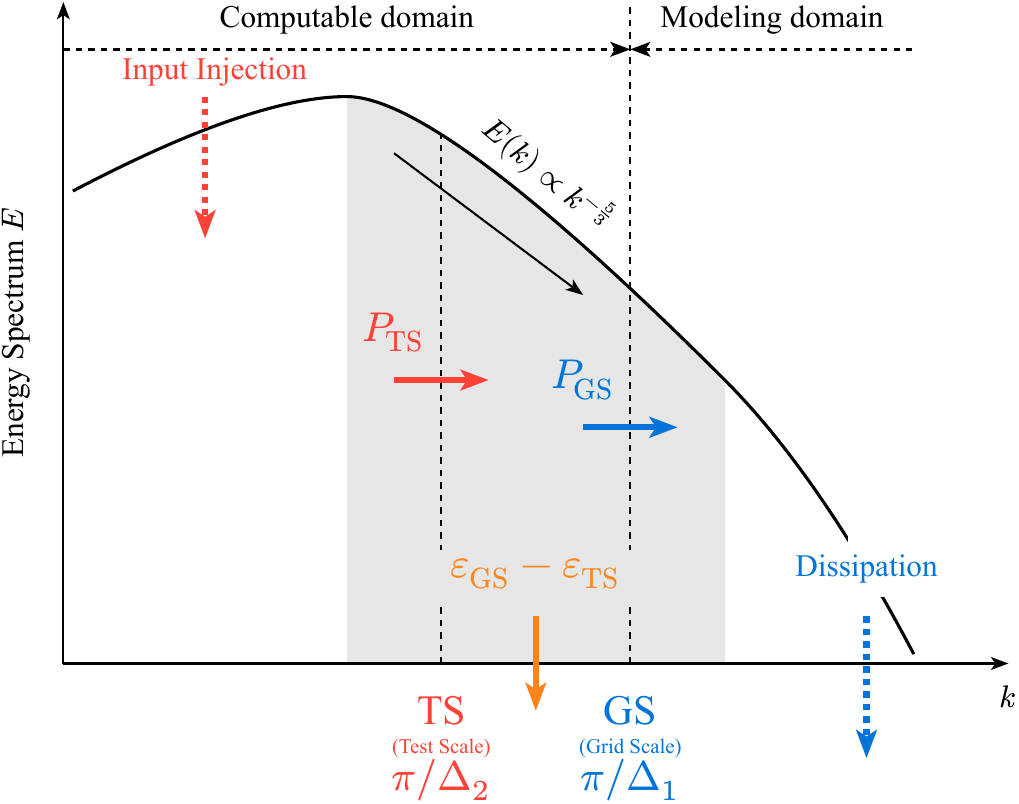}
    \caption{
        Schematic of the energy budget in scale space.
        Energy production (energy flux from large to small scales) at the test scale $\Delta_2$ and the grid scale $\Delta_1$ are denoted by $P_\mathrm{TS}$ and $P_\mathrm{GS}$, respectively, where the corresponding wavenumber $k$ is related to the physical scale $\ell$ by $\ell = \pi/k$.
        The energy dissipated between these two scales equals $\varepsilon_\mathrm{TS} - \varepsilon_\mathrm{GS}$, where $\varepsilon_\mathrm{TS}$ and $\varepsilon_\mathrm{GS}$ are the energy dissipation rates up to the test and grid scales, respectively (see Eq.~\eqref{eq:local_equilibrium_1}).
    }
    \label{fig:energy_spectral_ipsgs}
\end{figure}

Here, we determine the time-dependent parameter $C_i(t)$ from a concept of ``information preservation.''
To this end, we introduce a Test Scale (TS) $\Delta_2$, which is larger than the GS ($\Delta_2 > \Delta_1$) as shown in Fig.~\ref{fig:energy_spectral_ipsgs}.
Note that in Sec.~\ref{subsec:LES_formulation}, we set the filter width $\Delta = \Delta_1$.
By denoting the filtering operation at the GS and TS by $\overline{(\cdot)}$ and $\widetilde{(\cdot)}$, respectively, we define the kinetic energy associated with these scales as
\begin{equation}
    k_\mathrm{GS} \coloneq \frac{\bar{u}_i\bar{u}_i}{2} \quad \text{and} \quad
    k_\mathrm{TS} \coloneq \frac{\widetilde{\bar{u}}_i\widetilde{\bar{u}}_i}{2}.
\end{equation}
Note that the TS velocity $\widetilde{\bar{u}}_i$ is defined as the GS velocity further filtered by the TS~\cite{Germano:1991}.

We further define the Test-to-Grid (TG) scale $\ell \in [\Delta_1, \Delta_2]$, which is available in the LES framework and contains useful information regarding the energy transfer mechanisms in the GS~\cite{Lozano:2022}.
We can define the TG kinetic energy as
\begin{equation}
    k_\mathrm{TG} \coloneq \widetilde{k_\mathrm{GS}} - k_\mathrm{TS},
\end{equation}
which evolves as
\begin{equation}
    \frac{\partial k_{\mathrm{TG}}}{\partial t} + \widetilde{\overline{u}}_j \frac{\partial k_{\mathrm{TG}}}{\partial x_j}
    = - \left( \widetilde{P_{\mathrm{GS}}} - P_{\mathrm{TS}} \right)
    - \left( \widetilde{\varepsilon_{\mathrm{GS}}} - \varepsilon_{\mathrm{TS}} \right)
    - \left( \widetilde{J_{\mathrm{GS}}} - J_{\mathrm{TS}} \right).
    \label{eq:tg_energy_equation}
\end{equation}
Here, $P$, $\varepsilon$, and $J$ denote the energy production, dissipation, and diffusion, respectively.
These quantities at the GS and TS are defined as
\begin{align}
    P_\mathrm{GS}           & \coloneq - \tau_{ij}\bar{S}_{ij}, \label{eq:energy_production}                                                                                                                                                                                                                                                                                                \\
    P_\mathrm{TS}           & \coloneq - \left(\widetilde{\overline{u_i u_j}} - \widetilde{\bar{u}}_i\widetilde{\bar{u}}_j\right) \widetilde{\bar{S}}_{ij}, \label{eq:ts_energy_production}                                                                                                                                                                                                 \\
    \varepsilon_\mathrm{GS} & \coloneq 2\nu \bar{S}_{ij} \bar{S}_{ij}, \label{eq:gs_energy_dissipation}                                                                                                                                                                                                                                                                                     \\
    \varepsilon_\mathrm{TS} & \coloneq 2\nu \widetilde{\bar{S}}_{ij} \widetilde{\bar{S}}_{ij}, \label{eq:ts_energy_dissipation}                                                                                                                                                                                                                                                             \\
    J_\mathrm{GS}           & \coloneq \frac{\partial}{\partial x_j} \left(\bar{u}_i\tau_{ij} + \bar{u}_j k_\mathrm{GS} + \frac{\bar{u}_j\bar{\Pi}}{\rho} - \nu\frac{\partial k_\mathrm{GS}}{\partial x_j}\right), \label{eq:gs_energy_transport}                                                                                                                                           \\
    J_\mathrm{TS}           & \coloneq \frac{\partial}{\partial x_j} \left(\widetilde{\bar{u}}_i\left(\widetilde{\overline{u_i u_j}} - \widetilde{\bar{u}}_i\widetilde{\bar{u}}_j\right) + \widetilde{\bar{u}}_j \widetilde{k_\mathrm{GS}} + \frac{\widetilde{\bar{u}}_j\widetilde{\bar{\Pi}}}{\rho} - \nu\frac{\partial k_\mathrm{TS}}{\partial x_j}\right).\label{eq:ts_energy_transport}
\end{align}

Spatial average of Eq.~\eqref{eq:tg_energy_equation} reads
\begin{equation}
    \left\langle\frac{\partial k_\mathrm{TG}}{\partial t} + \widetilde{\bar{u}}_j\frac{\partial k_\mathrm{TG}}{\partial x_j}\right\rangle_V
    = \left\langle
    - \left( \widetilde{P_{\mathrm{GS}}} - P_{\mathrm{TS}} \right)
    - \left( \widetilde{\varepsilon_{\mathrm{GS}}} - \varepsilon_{\mathrm{TS}} \right)
    - \left( \widetilde{J_{\mathrm{GS}}} - J_{\mathrm{TS}} \right)
    \right\rangle_V.
\end{equation}
Here, $\langle \cdot \rangle_V$ denotes the spatial integral $\int_V \cdot \,\mathrm{d}V$.
Under the periodic boundary condition, the advection and diffusion terms vanish to arrive at
\begin{equation}
    \left\langle\frac{\partial k_\mathrm{TG}}{\partial t} \right\rangle_V
    = \left\langle
    - \left( \widetilde{P_{\mathrm{GS}}} - P_{\mathrm{TS}} \right)
    - \left( \widetilde{\varepsilon_{\mathrm{GS}}} - \varepsilon_{\mathrm{TS}} \right)
    \right\rangle_V.
\end{equation}
Then, we assume that the temporal variation of $k_\mathrm{TG}$ is insensitive to the energy production and dissipation terms
\begin{equation}
    \left\langle\left|\frac{\partial k_\mathrm{TG}}{\partial t}\right|\right\rangle_V
    \ll \left\langle\left|\widetilde{P_{\mathrm{GS}}} - P_{\mathrm{TS}}\right|\right\rangle_V, \left\langle\left|\widetilde{\varepsilon_{\mathrm{GS}}} - \varepsilon_{\mathrm{TS}}\right|\right\rangle_V,
    \label{eq:tg_energy_equation_periodic}
\end{equation}
to obtain the local inter-scale equilibrium~\cite{murota:2003}
\begin{equation}
    \left\langle \widetilde{P_{\mathrm{GS}}} + \varepsilon_{\mathrm{GS}}\right\rangle_V
    = \left\langle \widetilde{P_{\mathrm{TS}}} + \varepsilon_{\mathrm{TS}} \right\rangle_V.
    \label{eq:local_equilibrium_1}
\end{equation}
Here, we assume that the local equilibrium relation~\eqref{eq:local_equilibrium_1} also holds within a spatially local subdomain $v \in V$ as
\begin{equation}
    \left\langle \widetilde{P_{\mathrm{GS}}} + \varepsilon_{\mathrm{GS}}\right\rangle_V
    = \left\langle \widetilde{P_{\mathrm{TS}}} + \varepsilon_{\mathrm{TS}} \right\rangle_V.
    \label{eq:local_equilibrium}
\end{equation}
Although this assumption does not hold rigorously, we show in Appendix~\ref{sec:Appendix_B} that it is approximately satisfied with $\sqrt[3]{v}$ larger than $\sim 10\Delta_1$.
By substituting the definitions~(\ref{eq:energy_production}--\ref{eq:ts_energy_dissipation}) into Eq.~\eqref{eq:local_equilibrium} and rearrange the terms including the SGS stress $\tau_{ij}$, we obtain
\begin{align}
    \Gamma_1            & = \Gamma_2 (C_1, C_4),                                                                                                                                                                                                                      \\
    \Gamma_1            & = \left\langle -\left(\widetilde{\bar{u}_i\bar{u}_j} - \widetilde{\bar{u}}_i\widetilde{\bar{u}}_j\right)\widetilde{\bar{S}}_{ij} + 2 \nu\widetilde{\bar{S}}_{ij}\widetilde{\bar{S}}_{ij} \right\rangle_v, \label{eq:local_equilibrium_left} \\
    \Gamma_2 (C_1, C_4) & = \left\langle \widetilde{\tau}_{ij}\widetilde{\bar{S}}_{ij} - \widetilde{\tau_{ij}\bar{S}_{ij}} + 2\nu\widetilde{\bar{S}_{ij}\bar{S}_{ij}} \right\rangle_v. \label{eq:local_equilibrium_right}
\end{align}
Note that we omit the time dependence $C_i(t)$ for notation simplicity.
Here, $\Gamma_2(C_1, C_4)$ contains $\tau_{ij}$ to be modeled, and thus depend on the model parameters $C_1$ and $C_4$ (see Eq.~\eqref{eq:selected_model}).
We formulate a maximization problem of the mutual information
\begin{equation}
    C_1, C_4 = \arg\max_{C_1, C_4} I(\Gamma_1 \colon \Gamma_2(C_1, C_4)),
\end{equation}
which states that the information content of the net energy fluxes at the test and grid scales are equivalent.
Note that Eq.~\eqref{eq:local_equilibrium_right} contains the GS dissipation term $2\nu\widetilde{\bar{S}_{ij}\bar{S}_{ij}}$, which is independent of $\tau_{ij}$, to avoid the indeterminacy of the maximization problem.

\begin{figure}[tb]
    \centering
    \includegraphics[width=0.7\linewidth]{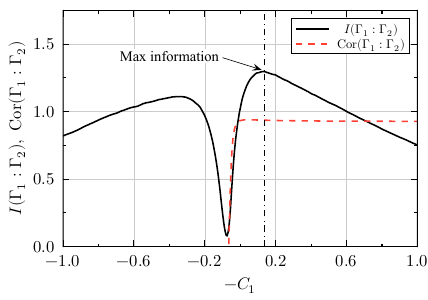}
    \caption{
        The model parameter $C_1$ dependency of the mutual information (solid black line) and correlation coefficient (red dashed line) between $\Gamma_1$ (computed from the DNS data) and $\Gamma_2(C_1, C_4)$ (including the SGS model) of a snapshot.
        Note that the other parameter $C_4$ is fixed to zero.
        The vertical black dash-dotted line indicates the maximum of the mutual information at $-C_1 = 0.135$.
    }
    \label{fig:estimation_parameter}
\end{figure}

We perform an \textit{a priori} test of the parameter estimation procedure using the DNS data of the periodic-box turbulence introduced in Sec.~\ref{subsec:Model_term_selection_via_a_priori_test}.
Figure~\ref{fig:estimation_parameter} compares $C_1$ dependency of the mutual information $I(\Gamma_1 \colon \Gamma_2(C_1, C_4))$ and the correlation coefficient $\mathrm{Cor}(\Gamma_1, \Gamma_2(C_1, C_4))$.
We find the maximum value of the mutual information at $C_1 = -0.135$, which is close to the empirically determined value of $-0.1$ reported in~\cite{Kobayashi:2005}.
However, the correlation coefficient does not show a clear maximum.
Note that we obtain $C_4 = -0.31$ using a similar approach.
Overall, the information-theoretic parameter estimation method appears to offer an advantage.

\begin{figure}[tb]
    \centering
    \includegraphics[width=0.8\linewidth]{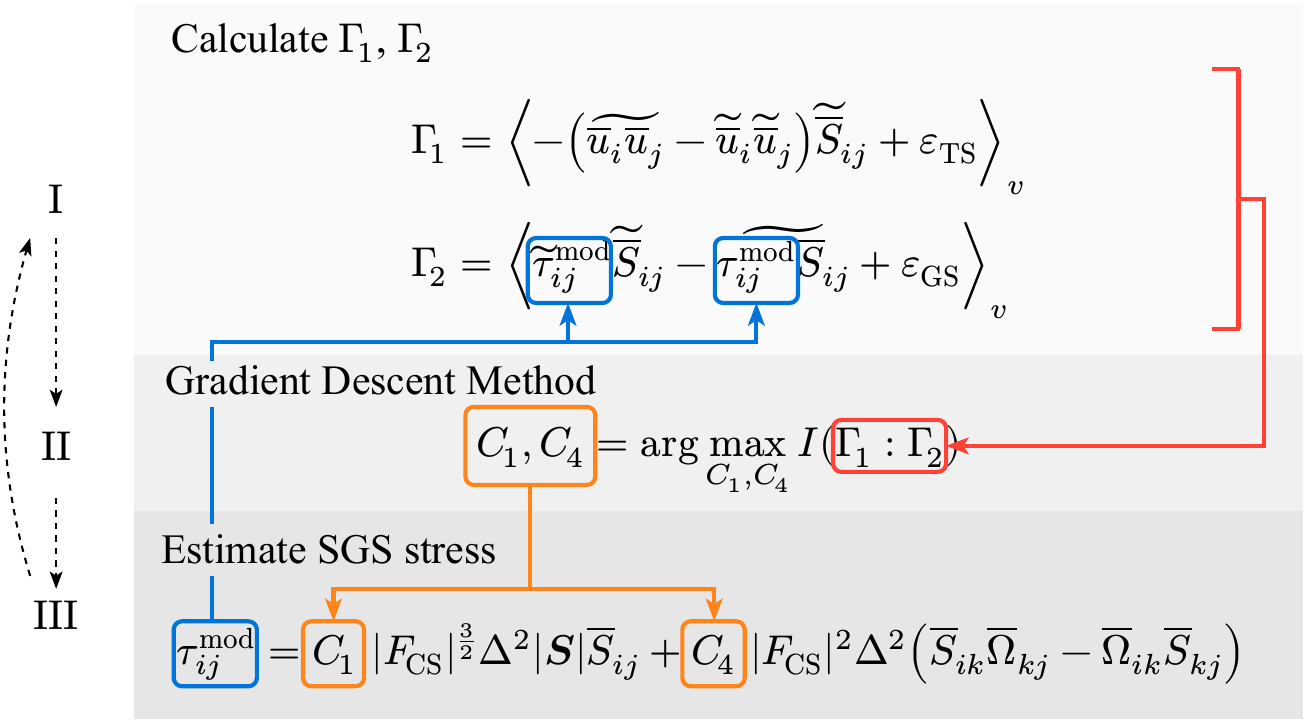}
    \caption{
        Step-by-step calculation procedure of the IP-CSM.
        We (I) evaluate $\Gamma_1$ and $\Gamma_2(C_1, C_4)$ from Eqs.~\eqref{eq:local_equilibrium_left} and \eqref{eq:local_equilibrium_right}, then (II) estimate the parameters $C_1$ and $C_4$ by maximizing the mutual information with the gradient descent method, and then (III) estimate the SGS stress.
    }
    \label{fig:step-by-step_IPCSM}
\end{figure}

Figure~\ref{fig:step-by-step_IPCSM} summarises the $\tau_{ij}^{\mathrm{mod}}$ estimation procedure of the proposed model, hereafter called as the Information-Preserving coherent structure Model (IP-CSM).
Before concluding this section, we stress the difference between our IP-CSM and its predecessor, the IP-SGS model proposed in~\cite{Lozano:2022}.
First, we employ the coherent structure function $F_{\mathrm{CS}}$ to construct the invariant model parameter $f_i(\boldsymbol{x}, t)$ (see Eq.~\eqref{eq:coherent_structure_function}) so that the model can be applied to the wall-bounded flows.
In contrast, the IP-SGS model does not explicitly consider this parameter and requires a damping function to ensure the near-wall behavior.
Second, our model estimate the model parameter $C_i(t)$ based on the nature of the scale-local equilibrium while the IP-SGS relies on the nontrivial scaling of the energy flux probability density function (PDF) obtained from DNS data.
Our approach is independent from the scaling laws extracted from DNS data.
Moreover, we measure the information preservation with the symmetric mutual information $I(X:Y)$ while the previous model employed the Kullback--Leibler (KL) divergence, which is not a appropriate measure of distance since $\mathrm{KL}(X, Y) \neq \mathrm{KL}(Y, X)$.

\section{A Posteriori test of the IP-CSM}
\label{sec:A_Posteriori_test_of_the_IP-CSM}

In this section, we conduct an \textit{a posteriori} test of the proposed IP-CSM in two flow configurations: periodic box and wall-bounded channel.
To evaluate the model performance, we compare our model with the Smagorinsky model~\cite{Smagorinsky:1963} and the coherent structure model~\cite{Kobayashi:2005}.

\subsection{Periodic box turbulence}
\label{subsec:Periodic_box_turbulence}

In this subsection, we compare the results of the DNS and each of the SGS models in the periodic box configuration.
Numerical conditions are summarised in Table~\ref{tab:periodic_box_turbulence}.
The reference SGS models are the Smagorinsky model~\cite{Smagorinsky:1963}
\begin{equation}
    \tau_{ij}^{\mathrm{mod}} = C_1\Delta^2 |\boldsymbol{S}| \bar{S}_{ij} \quad \text{with} \quad C_1 = -2\cdot 0.17^2,
\end{equation}
and the coherent structure model~\cite{Kobayashi:2005}
\begin{equation}
    \tau_{ij}^{\mathrm{mod}} = C_1 f_1\Delta^2 |\boldsymbol{S}| \bar{S}_{ij}, \quad \text{with} \quad C_1 = -0.1 \quad \text{and} \quad f_1 = |F_{\mathrm{CS}}|^{\frac{3}{2}}.
\end{equation}
In the IP-CSM calculation, the size of the local subdomain is set to $v/\Delta^3 = 8$, where $\Delta$ denotes the GS.
To reduce computational cost, the parameters $C_1$ and $C_4$ are estimated and updated every 100 time steps.
Details of these numerical settings are provided in Appendices~\ref{sec:Appendix_B} and \ref{sec:Appendix_C}.
For reference, a posteriori test showed that the average and standard deviation of the parameters are $C_1= -0.21 \pm 0.11$ and $C_4 = -0.32 \pm 0.26$.

\begin{figure}[tb]
    \centering
    \begin{minipage}{0.45\textwidth}
        \centering
        \includegraphics[width=\textwidth]{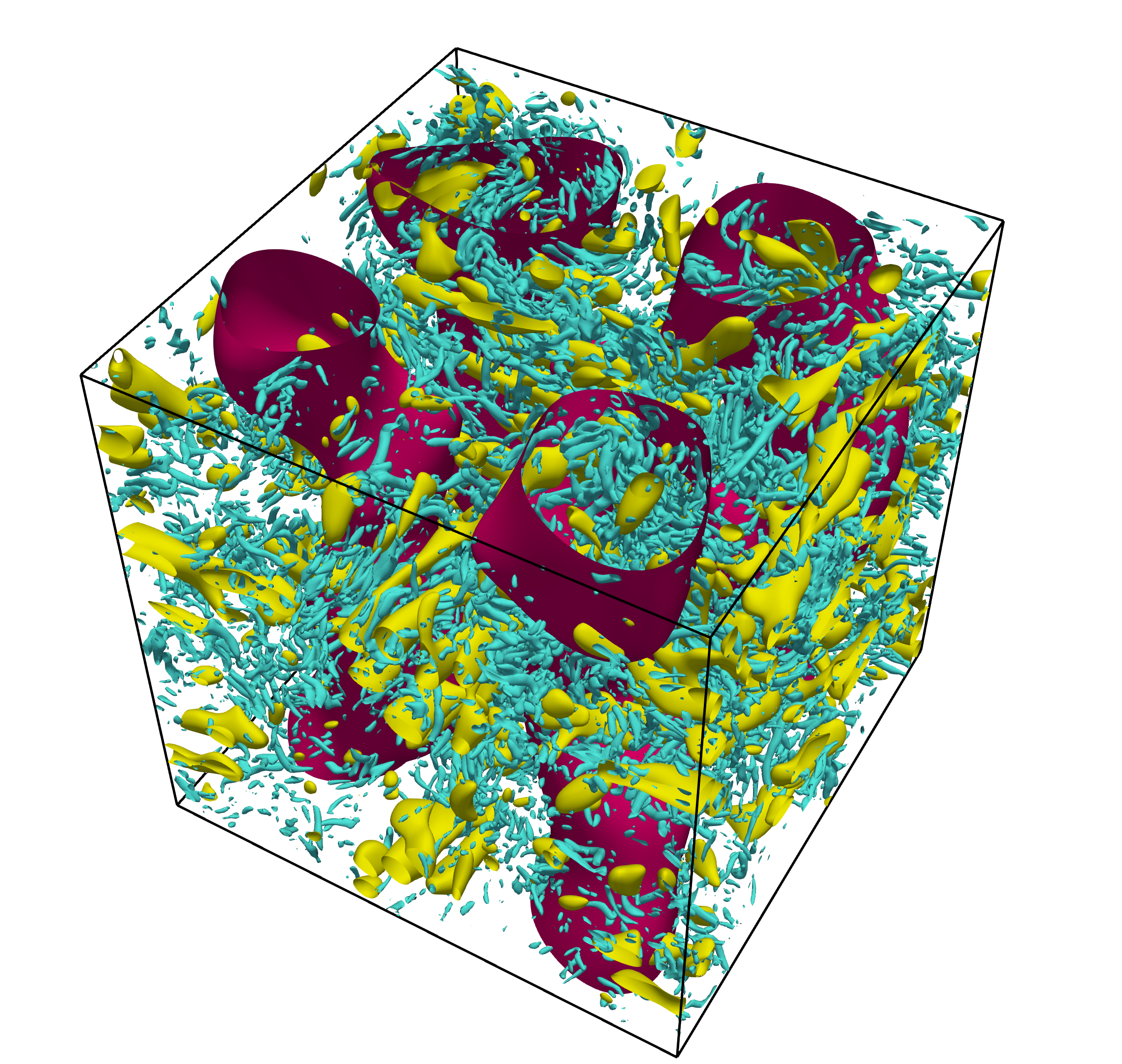}
        \subcaption{DNS}
    \end{minipage}
    \begin{minipage}{0.45\textwidth}
        \centering
        \includegraphics[width=\textwidth]{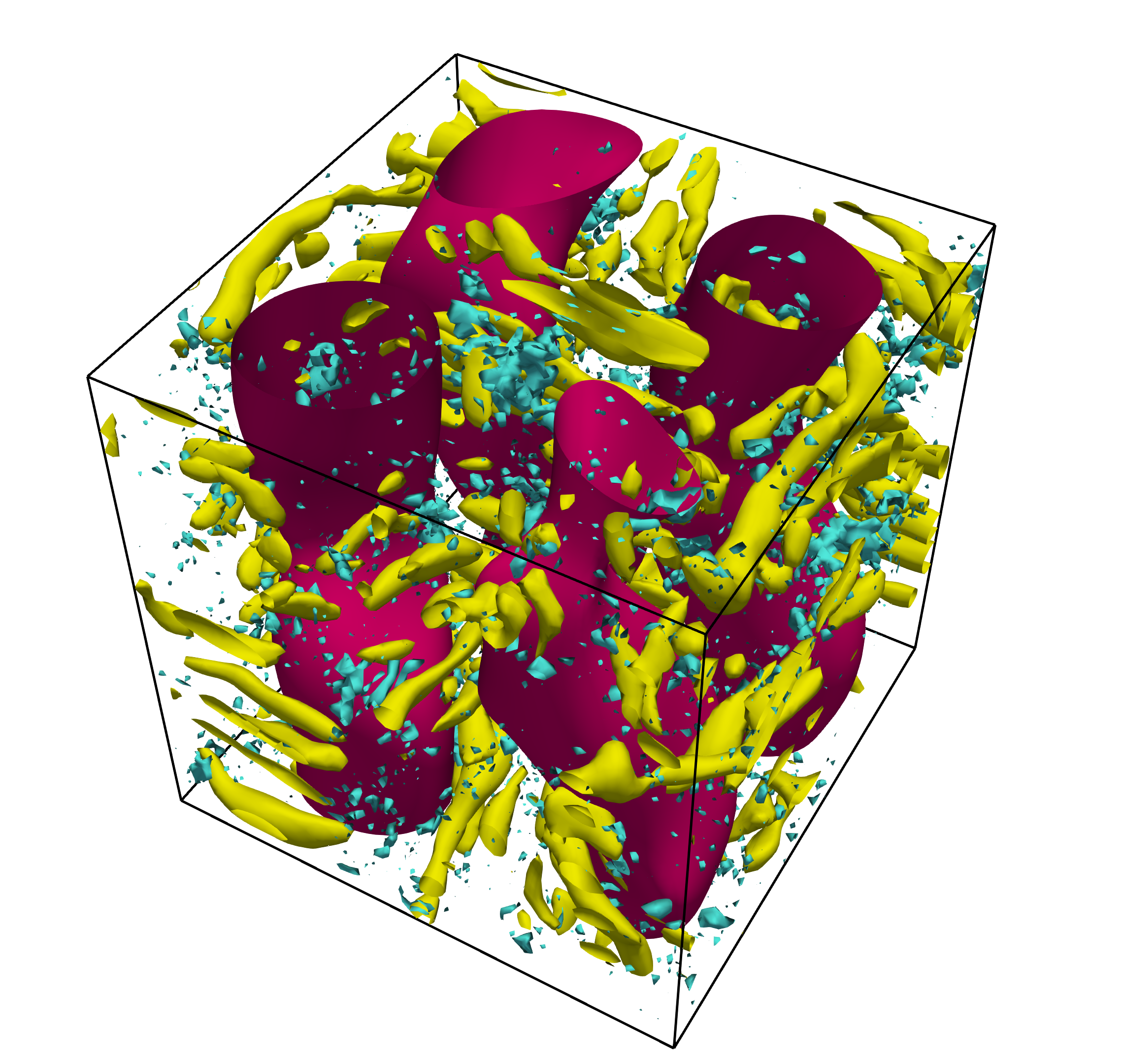}
        \subcaption{IP-CSM}
    \end{minipage}
    \caption{
        Isosurfaces of the enstrophy field in periodic-box turbulence computed by (a) DNS and (b) IP-CSM.
        For each case, we extract the hierarchical vortex structures using a band-pass filter in the range $k_c/\sqrt{2} \leq k \leq k_c\sqrt{2}$, where $k_c$ is the cutoff wavenumber.
        The red, yellow, and cyan vortex structures correspond to $k_c =2, 4$, and $8$, respectively.
    }
    \label{fig:quasi_periodic_vorticity}
\end{figure}

First, we compare the hierarchical vortex structures generated by the DNS and IP-CSM, as shown in Fig.~\ref{fig:quasi_periodic_vorticity}.
The DNS snapshot shows the four large-scale vortices directly driven by the Taylor--Green forcing (Eq.~\eqref{eq:Taylor--Green_force}) and the hierarchical vortex structures generated by the scale-local energy cascade process~\cite{Goto:2017}.
The IP-CSM snapshot reproduces similar coherent structures, but the small-scale structures are less pronounced than the DNS due to its small resolution.

\begin{figure}[tb]
    \centering
    \includegraphics[width=0.7\textwidth]{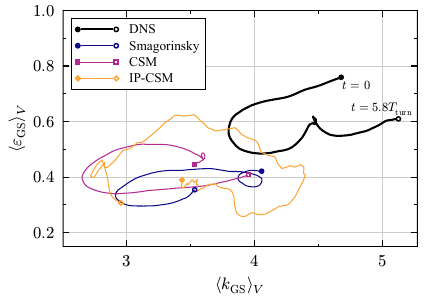}
    \caption{
        Quasi-cyclic temporal fluctuations in periodic box turbulence.
        The horizontal and vertical axes represent the volume-averaged kinetic energy $\langle k_\mathrm{GS}\rangle_V(t)$ and energy dissipation rate $\langle \varepsilon_\mathrm{GS}\rangle_V(t)$ of the GS, respectively.
        Different data sets show DNS (black), Smagorinsky model (blue), coherent structure model (red), and information-preserving coherent structure model (yellow).
        Note that the DNS results are computed from the coarse-grained velocity field at scale $\Delta_1$.
        Filled and open symbols represent the time $t = 0$ and $t = 5.8T_\mathrm{turn}$, respectively.
    }
    \label{fig:quasi_periodic_diagram}
\end{figure}

Next, we compare the quasi-cyclic temporal fluctuations observed in turbulence driven by the steady forcing~\cite{Goto:2017}.
Figure~\ref{fig:quasi_periodic_diagram} shows the GS energy cycle between $\langle k_\mathrm{GS}\rangle_V(t)$ and $\langle \varepsilon_\mathrm{GS}\rangle_V(t)$.
The DNS time series display a counter-clockwise trajectory, a characteristic of the energy cascade~\cite{Goto:2017}.
The SGS models reproduce similar cycles, while the IP-CSM exhibits intense fluctuations compare to the others due to its dynamic formulation.

\begin{figure}[tb]
    \centering
    \begin{minipage}{0.45\textwidth}
        \centering
        \includegraphics[width=\textwidth]{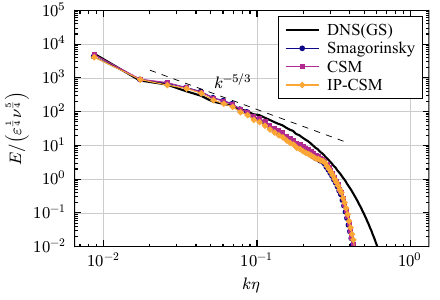}
        \subcaption{}
    \end{minipage}
    \begin{minipage}{0.45\textwidth}
        \centering
        \includegraphics[width=\textwidth]{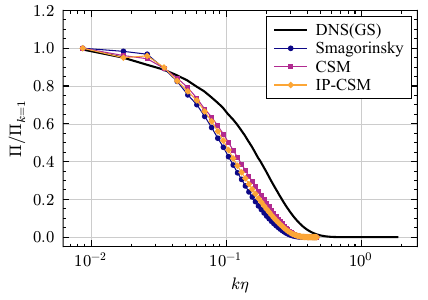}
        \subcaption{}
    \end{minipage}
    \caption{
        (a) Energy and (b) energy flux spectra in the periodic-box turbulence.
        In panel (a), the dashed line indicates the $k^{-5/3}$ scaling.
        Note that the DNS (GS) results are computed from the coarse-grained velocity field at scale $\Delta_1$.
    }
    \label{fig:periodic_spectrum}
\end{figure}

Next, we compare the turbulence statistics.
Figure~\ref{fig:periodic_spectrum} shows the time-averaged energy spectrum $E(k)$ and energy flux spectrum $\Pi (k)$.
In the energy spectrum, all of the SGS models reproduce the scaling with an exponent close to $k^{-5/3}$, a defining characteristic of developed turbulence.
The energy flux spectra also exhibit similar scaling across different models.
These results indicate that the IP-CSM reproduces the second-order statistics in the periodic box turbulence.

\begin{figure}[tb]
    \centering
    \includegraphics[width=0.7\textwidth]{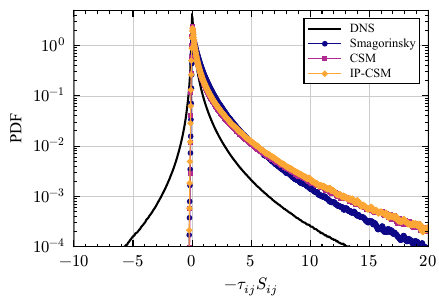}
    \caption{
        Probability density function of the energy production in the periodic box turbulence shown in semi-log plot.
        The SGS stress $\tau_{ij}$ is evaluated according to the definition of each model.
    }
    \label{fig:energy_pdf}
\end{figure}

Finally, we show the PDF of the energy production $-\tau_{ij}\bar{S}_{ij}$ in Fig.~\ref{fig:energy_pdf}.
Compared with the DNS, the SGS models are characterized by the absence of the negative energy production, known as the energy backscatter.
For the IP-CSM (Eq.~\eqref{eq:selected_model}), $M_{ij}^{[1]}$ always produces positive energy when $C_1 < 0$ and $M_{ij}^{[4]}$ does not contribute to the energy production.
Thus, the IP-CSM reproduces the PDF similar to the other eddy-viscosity models.
This characteristics is supported by a recent study~\cite{Vela:2022} suggesting that the reproduction of the backscatter in an SGS model is not necessarily essential; rather, accurate prediction the net energy production is more important.

\subsection{Wall-bounded turbulence}
\label{subsec:Wall-bounded_turbulence}

\begin{table}[tb]
    \centering
    \caption{
        Parameters and statistical quantities of the channel turbulence.
        The grid resolution $N$;
        the friction Reynolds number $\mathrm{Re}_\tau = u_\tau\delta/\nu$, where $u_\tau = \sqrt{\tau_w/\rho}$ is the friction velocity based on the wall shear stress $\tau_w$ and $\delta$ is the channel half-width;
        the computational domain size $L$; and
        the spatial resolution $\Delta^+ = \Delta / (\nu/u_\tau)$.
    }
    \begin{tabular}{ccccccc} \hline\hline
                               & $N_x\times N_y\times N_z$  & $\mathrm{Re}_\tau$   & $L_x\times L_y\times L_z$                  & $\Delta x^+ $ & $\Delta y^+$    & $\Delta z^+$ \\ \hline
        DNS~\cite{Kozuka:2009} & $2048\times 480\times 512$ & \multirow{2}{*}{395} & \multirow{2}{*}{$6.4\times 2.0\times 3.2$} & $1.23$        & $0.111$--$2.13$ & $2.47$       \\
        LES                    & $64\times 64\times 64$     &                      &                                            & $39.5$        & $1.23$--$28.89$ & $19.75$      \\ \hline
    \end{tabular}
    \label{tab:channel_conditions}
\end{table}

In this subsection, we examine the near-wall behavior of the IP-CSM in fully developed channel turbulence driven by a pressure gradient.
For the numerical simulations, we implemented the IP-CSM to the well-validated in-house DNS solver used in previous studies~\cite{Abe:2001, Tsukahara:2005, Kozuka:2009} and incorporated each SGS model into it.
Table~\ref{tab:channel_conditions} summarises the parameters and statistical quantities of the channel turbulence.
To resolve the near-wall region, we employ a non-uniform grid in the wall-normal $y$ direction.

For comparison, we again employ the Smagorinsky model and the coherent structure model.
Since the standard Smagorinsky model does not satisfy the near-wall asymptotic condition, we introduce the van Driest damping function~\cite{Van:1956}
\begin{equation}
    f_s = 1 - \exp\left(-\frac{y^+}{A^+}\right) \quad \text{with} \quad A^+ = 25,
    \label{eq:van_driest_damping}
\end{equation}
to redefine the SGS stress
\begin{equation}
    \tau_{ij}^{\mathrm{mod}} = C_1 \left(f_s \Delta\right)^2 |\boldsymbol{S}| \bar{S}_{ij} \quad \text{with} \quad C_1 = -2\cdot 0.1^2.
\end{equation}
Here, the superscript $(\cdot)^+$ denotes the normalization by the friction velocity $u_\tau$ and the viscous length scale $\nu/u_\tau$.
For reference, a posteriori test of the proposed IP-CSM yielded $C_1 = -0.24 \pm 0.09$ and $C_4 = -0.064 \pm 0.178$ in the channel turbulence.

\begin{figure}[tb]
    \centering
    \begin{minipage}{0.45\textwidth}
        \centering
        \includegraphics[width=\textwidth]{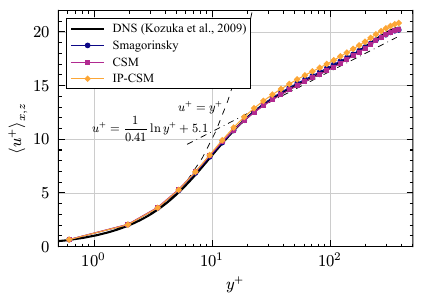}
        \subcaption{}
    \end{minipage}
    \begin{minipage}{0.45\textwidth}
        \centering
        \includegraphics[width=\textwidth]{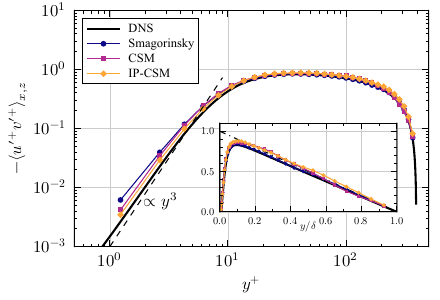}
        \subcaption{}
    \end{minipage}
    \caption{
            (a) Mean streamwise velocity $\langle u^+\rangle_{x, z}$ and (b) Reynolds shear stress $-\langle u'^+ v'^+ \rangle_{x, z} = - \langle \overline{u}'^+ \overline{v}'^+ \rangle_{x, z} - \langle\tau_{xy}^{\mathrm{mod}}\rangle_{x, z}$ profile as a function of the wall-normal distance $y^+$ in the channel.
            The dashed line in (a) indicates the viscous sublayer scaling $u^+ = y^+$, and the dash--dotted line represents the logarithmic law $u^+ = (1/0.41)\ln y^+ + 5.1$.
            The dashed line in (b) indicates the $y^3$ scaling, and the inset shows the same profile in linear-linear axes where the dash-dotted line represents the total shear stress, $1 - y/\delta$.
    }
    \label{fig:channel_wall_and_channel_stress}
\end{figure}

First, we show the mean streamwise velocity profile in Fig.~\ref{fig:channel_wall_and_channel_stress}~(a).
The viscous sublayer and log-law scaling observed in the DNS are well reproduced by the IP-CSM and comparable to the conventional SGS models.

Next, we show the Reynolds shear stress profile in Fig.~\ref{fig:channel_wall_and_channel_stress}~(b).
The DNS exhibits the $y^3$ scaling in the near-wall region.
Both the Smagorinsky model associated with the van Driest damping function of Eq.~\eqref{eq:van_driest_damping} and the coherent structure model reproduce the same $y^3$ scaling by the incompressibility condition.
The IP-CSM also reproduces the scaling, owing to the invariant parameters $f_i(\bm{x}, t)$ incorporating the coherent structure function.

\begin{figure}[tb]
    \centering
    \includegraphics[width=0.7\textwidth]{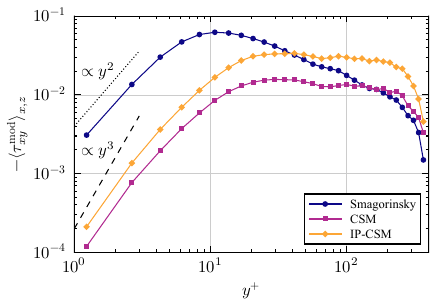}
    \caption{
        SGS stress $-\langle\tau_{xy}\rangle_{x, z}$ as a function of the wall-normal distance $y^+$ in the channel turbulence.
        The SGS stress $\tau_{xy}$ is evaluated according to the definition of each model.
        The dashed line indicates the $y^3$ scaling, and the dotted line represents the $y^2$ scaling.
    }
    \label{fig:channel_sgs_stress}
\end{figure}

Third, we show the SGS stress profile in Fig.~\ref{fig:channel_sgs_stress}.
The Smagorinsky model is associated with the $y^2$ scaling, which comes from the van Driest damping function~\cite{Chen:2025}.
By contrast, the CSM and IP-CSM reproduce a scaling close to $y^3$ as they satisfy the near-wall asymptotic condition for the SGS stress, following our formulation in \S~\ref{subsec:Invariant_parameter_determination_using_coherent_structure_function}.

\begin{figure}[tb]
    \centering
    \begin{minipage}{0.45\textwidth}
        \centering
        \includegraphics[width=\textwidth]{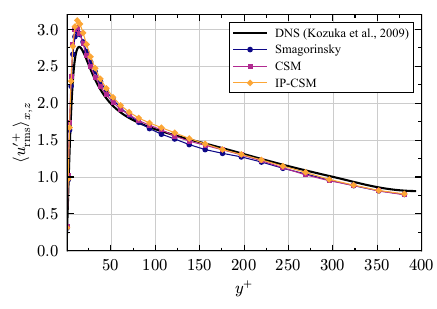}
        \subcaption{}
    \end{minipage}
    \begin{minipage}{0.45\textwidth}
        \centering
        \includegraphics[width=\textwidth]{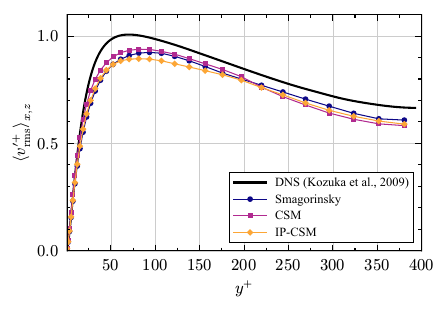}
        \subcaption{}
    \end{minipage}
    \begin{minipage}{0.45\textwidth}
        \centering
        \includegraphics[width=\textwidth]{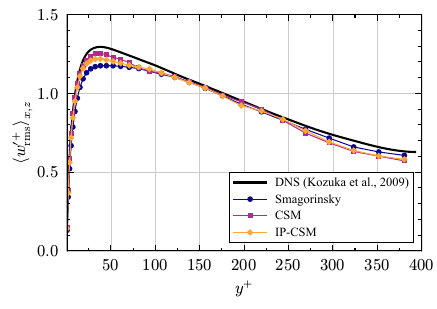}
        \subcaption{}
    \end{minipage}
    \caption{
        Root-mean-square velocity fluctuations profile in channel turbulence in its (a) streamwise $x$, (b) wall-normal $y$, and (c) spanwise $z$ components.
    }
    \label{fig:channel_rms}
\end{figure}

Finally, we show the root-mean-square velocity fluctuations profile in Figure~\ref{fig:channel_rms} to assess the anisotropic characteristics of the channel turbulence.
Although we expected the nonlinear IP-CSM to reproduce the anisotropy, the results are more or less collapsed to the other SGS models.
Higher Reynolds numbers or coarser grid resolutions may trigger the model's anisotropic feature~\cite{Marstorp:2009}.

\section{Discussion and Conclusion}
\label{sec:Discussion_and_Conclusion}

Many conventional SGS models require model parameters to be determined \textit{a priori}.
One even has to trial and error different model parameters before obtaining reliable results.
The model parameters vary for different flow configurations---for example, $C_1\approx -2\cdot 0.17^2$ in periodic box turbulence~\cite{Lilly:1966} and $C_1\approx -2\cdot 0.1^2$ in wall-bounded turbulence~\cite{Morinishi:2001}.
Germano and Lilly~\cite{Germano:1991,Lilly:1992} proposed a dynamic algorithm to determine these parameters \textit{in situ}.
Although this approach has inspired many subsequent studies~\cite{Yuan:2022,Rozema:2022}, these methods suffer from numerical instability and limited applicability to various flow configurations.
Murota~\cite{murota:2003} proposed a dynamic parameter estimation based on the local equilibrium hypothesis, but it requires different modeling strategies at the GS and TS to avoid the indeterminacy.
As an alternative approach, Lozano-Durán and Arranz~\cite{Lozano:2022} proposed the IP-SGS model, which estimates the model parameters from the KL divergence between the inter-scale energy flux PDFs at two scales by taking advantage of the self-similar turbulent statistics.
Their model was validated for homogeneous isotropic turbulence, and later applied to compressible channel turbulence~\cite{Williams:2022} by incorporating a wall function.

In this study, we extended the IP-SGS formulation to propose the ``information-preserving coherent structure model'' (IP-CSM) by reinterpreting the local inter-scale equilibrium hypothesis from an information-theoretic perspective.
To estimate the model parameters, we consider the information preservation (or similarity between two PDFs) between two adjacent scales using mutual information.
Our model belongs to a data-driven framework as it dynamically determines time-varying parameters while remaining physically interpretable rather than behaving as a black-box model.
Furthermore, we employ the coherent structure function to construct the invariant parameter, which allows the model to inherently satisfy the near-wall asymptotic condition without wall function.
In comparison to the IP-SGS model~\cite{Lozano:2022}, our model employs more straightforward formulation with physically consistent nature.

We validated the proposed IP-CSM with two flow configurations.
In periodic box turbulence, the model reproduced some low-order statistics, including the energy spectrum with the $k^{-5/3}$ scaling, and its accuracy is comparable to the existing models.
In channel turbulence, the model also reproduced several key statistical quantities including the velocity and Reynolds stress profiles, with accuracy comparable to the existing models.
We stress here that our model successfully captured the near-wall behavior without wall function, by employing the coherent structure function.

Although not shown, the proposed model also reproduces the nature of decaying turbulence.
This capability clearly distinguishes it from the single-parameter global model~\cite{Park:2006} for the entire computational domain, which fails to reproduce decaying turbulence~\cite{Lee:2010}.

We address some perspectives of the IP-CSM.
First, our method employs the gradient descent method to estimate the model parameters (see Fig.~\ref{fig:step-by-step_IPCSM}), which requires high computational cost.
Industrial flow simulations with this model require intense reduction in its computational cost.
To this end, we investigated the impact of the parameter estimation interval on the model accuracy in Appendix~\ref{sec:Appendix_C}.
Second, our model indeed has a new hyperparameter: a local subdomain size, or the test scale, in which the local equilibrium hypothesis is evaluated (see Appendix~\ref{sec:Appendix_B}).
Moreover, the stability of the model remains unclear when we select a modeling term that reproduces the backscatter, as discussed in Fig.~\ref{fig:energy_pdf} and Appendix~\ref{sec:Appendix_A}.
Last, our method relies on the concept of information preservation between adjacent scales, but it may be better to consider information transfer (causality) in both forward~\cite{Tanogami:2024, Araki:2024} and backward~\cite{Vela-Martín:2025, Freitas:2026} directions, rather than information preservation.

\begin{acknowledgments}
    The authors thank Drs. Hiromichi~Kobayashi and Kazuhiro~Inagaki for their helpful comments and feedback.
    T.~T. was supported by JSPS KAKENHI, Grant-in-Aid for Scientific Research(S), Grant Numbers 21H05007 and 26K21736.
    R.~A. was supported by JSPS KAKENHI, Grant-in-Aid for Research Activity Start-up, Grant Number JP24K22942 and JST PRESTO, Japan, Grant Number JPMJPR25K1.
    For the purpose of Open Access, a CC-BY public copyright licence has been applied by the authors to the present document and will be applied to all subsequent versions up to the Author Accepted Manuscript arising from this submission.
\end{acknowledgments}

\appendix
\section{A Case with Three Model Terms}
\label{sec:Appendix_A}

\begin{figure}[tb]
    \centering
    \includegraphics[width=0.95\linewidth]{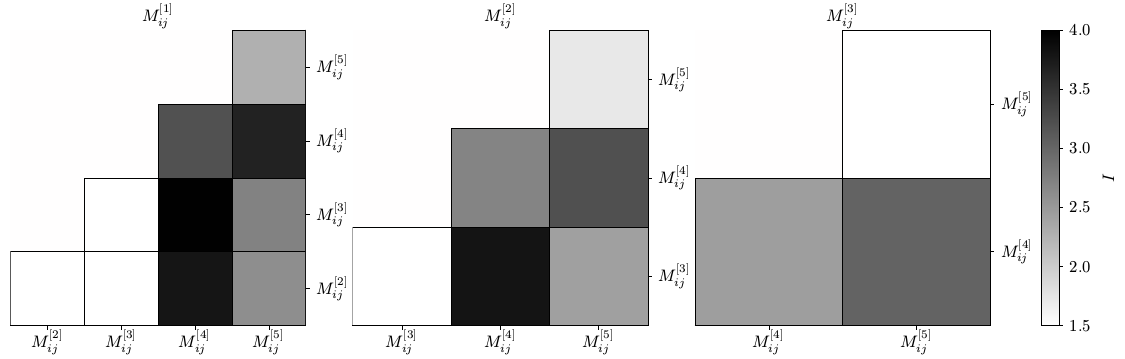}
    \caption{
        Mutual information between the SGS stress computed directly from DNS and the SGS stress modeled by selecting three terms from five candidate model components.
        Note that we evaluate the force and energy flux defined in terms of the SGS stress (see Eq.~\eqref{eq:mi_sgs_force_energy}).
        The diagonal elements represent the mutual information obtained with the two terms.
        See Fig.~\ref{fig:mi_mix} for the same figure with two terms.
    }
    \label{fig:mi_mix3}
\end{figure}

In the main text, we discussed the model consists of two terms out of five candidates: see Eqs.~\eqref{eq:sgs_stress_model} and~\eqref{eq:selected_model}.
Here, we consider the model with three terms.
Figure~\ref{fig:mi_mix3} shows the mutual information between the SGS stress computed directly from DNS and the SGS stress modeled by the three terms selected from five candidates.
For the two-component combinations shown in Fig.~\ref{fig:mi_mix}, the pair $(M_{ij}^{[1]}, M_{ij}^{[4]})$ yielded the highest mutual information, and we observe a similar trend in this case.
Among all combinations, the set $(M_{ij}^{[1]}, M_{ij}^{[3]}, M_{ij}^{[4]})$ results in the highest mutual information while $(M_{ij}^{[1]}, M_{ij}^{[2]}, M_{ij}^{[4]})$ and $(M_{ij}^{[1]}, M_{ij}^{[4]}, M_{ij}^{[5]})$ exhibit similar values.
Note that backscatter will be incorporated by including the term $M_{ij}^{[2]}$ or $M_{ij}^{[3]}$, which may destabilize the model.
Also note that the $M_{ij}^{[5]}$ term, which does not contribute to the energy flux, produces one of the highest mutual information in the $(M_{ij}^{[1]}, M_{ij}^{[4]}, M_{ij}^{[5]})$ combination.
It is intriguing to compare this information-theoretic observation with a recent claim that the backscatter is unnecessary~\cite{Vela:2022}, but it is out of the scope of the present paper.

\section{Evaluation of Local Equilibrium Hypothesis in Physical Space}
\label{sec:Appendix_B}

In this section, we discuss the local equilibrium hypothesis in both wavenumber and physical space.
By assuming the periodic boundary condition, the energy equation in wavenumber space reads
\begin{equation}
    \frac{\partial E(k, t)}{\partial t} = T(k, t) - 2\nu k^2 E(k, t) + F(k, t).
\end{equation}
Here, $E(k, t)$ is the energy spectrum, $T(k, t)$ is the energy transfer spectrum, and $F(k, t)$ is the energy injection spectrum.
Integrating this equation between the GS $k_1 = \pi/\Delta_1$ and TS $k_2 = \pi/\Delta_2$ allows us to consider the energy balance at the intermediate scale (hereafter denoted by ``TG'') $K_\mathrm{TG}(t) \coloneq \int_{k_2}^{k_1} E(k', t)\,\mathrm{d}k'$, which evolves as
\begin{align}
    \frac{\partial K_\mathrm{TG}(t)}{\partial t} & = -\Pi_\mathrm{TG}(t) - \varepsilon_\mathrm{TG}(t) + F_\mathrm{TG}(t), \\
    \varepsilon_\mathrm{TG}(t)                   & \coloneq \int_{k_2}^{k_1} 2\nu k'^2 E(k', t)\,\mathrm{d}k',            \\
    \Pi_\mathrm{TG}(t)                           & \coloneq -\int_{k_2}^{k_1} T(k', t)\,\mathrm{d}k',                     \\
    F_\mathrm{TG}(t)                             & \coloneq \int_{k_2}^{k_1} F(k', t)\,\mathrm{d}k'.
\end{align}
Here, we impose two assumptions.
First, we assume that the external force is concentrated in the large scales $k \ll k_1$ so that $F_\mathrm{TG}(t) \approx 0$.
Next, we assume that the magnitude of the left-hand side is negligible compared to the right-hand side, or $|\partial_t K_\mathrm{TG}| \ll \Pi_\mathrm{TG}, \varepsilon_\mathrm{TG}$.
Then, we obtain the scale-local energy equilibrium balance~\cite{murota:2003}
\begin{equation}
    \Pi_\mathrm{TG}(t) + \varepsilon_\mathrm{TG}(t) \approx 0.
\end{equation}

We remark here that the physical quantities at the TG can be expressed by the difference between two low-pass quantities at the GS and TS,
\begin{align}
    \int_{k_2}^{k_1} \Pi(k') \mathrm{d}k'         & = \int_0^{k_1} \Pi(k') \,\mathrm{d}k' - \int_0^{k_2} \Pi(k') \,\mathrm{d}k',                 \\
    \Pi_\mathrm{TG}                               & = \Pi_\mathrm{GS} - \Pi_\mathrm{TS},                                                         \\
    \int_{k_2}^{k_1} \varepsilon(k') \mathrm{d}k' & = \int_0^{k_1} \varepsilon(k') \,\mathrm{d}k' - \int_0^{k_2} \varepsilon(k') \,\mathrm{d}k', \\
    \varepsilon_\mathrm{TG}                       & = \varepsilon_\mathrm{GS} - \varepsilon_\mathrm{TS}.
\end{align}
Also, note that we can rewrite the low pass filtering in wavenumber space by spatial integral of coarse-grained quantity in physical space
\begin{align}
    \int_0^{k_1} \Pi(k') \mathrm{d}k'         & = \frac{1}{V}\left\langle \Pi_\mathrm{GS} \right\rangle_V = \frac{1}{V}\left\langle\widetilde{\Pi_\mathrm{GS}} \right\rangle_V,\label{eq:lowpass_integral_conversion} \\
    \int_0^{k_2} \Pi(k') \mathrm{d}k'         & = \frac{1}{V}\left\langle \Pi_\mathrm{TS} \right\rangle_V,                                                                                                            \\
    \int_0^{k_1} \varepsilon(k') \mathrm{d}k' & = \frac{1}{V}\left\langle \varepsilon_\mathrm{GS} \right\rangle_V = \frac{1}{V}\left\langle\widetilde{\varepsilon_\mathrm{GS}} \right\rangle_V,                       \\
    \int_0^{k_2} \varepsilon(k') \mathrm{d}k' & = \frac{1}{V}\left\langle \varepsilon_\mathrm{TS} \right\rangle_V.
\end{align}
when the flow field is statistically homogeneous and isotropic.
Here, the last equality in Eq.~\eqref{eq:lowpass_integral_conversion} follows from the normality of the filter function ($\int_V G(\boldsymbol{x}) = 1$).
Now, we can express the local equilibrium balance in physical space as
\begin{equation}
    \frac{1}{V}\left\langle (\widetilde{\Pi_\mathrm{GS}} + \widetilde{\varepsilon_\mathrm{GS}}) - (\Pi_\mathrm{TS} + \varepsilon_\mathrm{TS}) \right\rangle_V \approx 0,
\end{equation}
where the argument $t$ is omitted to keep the equation simple.

\begin{figure}[tp]
    \centering
    \includegraphics[width=0.7\linewidth]{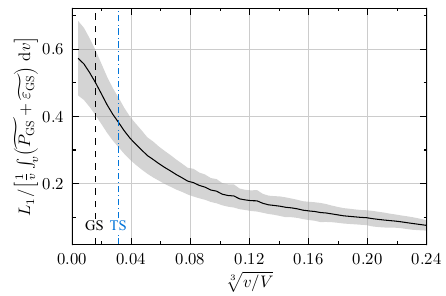}
    \caption{
        Error of the net energy flux $P + \varepsilon$ against the local domain size.
        The horizontal axis shows the scale ratio, defined by the cubic root of the local domain $v$ divided by the whole computational domain $V$.
        The error is normalized by the energy flux at the GS.
        The black dashed line and blue dotted line represent the GS and TS, respectively.
    }
    \label{fig:local_eq_norm}
\end{figure}

We assume the local equilibrium hypothesis to hold for a local subdomain $v \in V$ in physical space to claim Eq.~\eqref{eq:local_equilibrium}.
To evaluate its validity, we assess the L1 norm error
\begin{equation}
    L_1 = \left|\frac{1}{v}\left\langle\left(\widetilde{P_\mathrm{GS}} + \widetilde{\varepsilon_\mathrm{GS}}\right) - \left(P_\mathrm{TS} + \varepsilon_\mathrm{TS}\right)\right\rangle_v\right|,
\end{equation}
of the net energy flux at the GS and TS.
In Fig.~\ref{fig:local_eq_norm}, the error remains at large values of \qtyrange{40}{50}{\percent} near the GS and TS, but it decreases as the local subdomain volume increases.
Thus, the relation holds with relatively small error of \qtyrange{10}{20}{\percent} when we select the local subdomain to be 5--10 times larger than the GS.
Throughout this study, we fixed this newly introduced model parameter to be $v/\Delta^3 = 8$, where $\Delta$ denotes the GS.

\section{Trade-off between Parameter Estimation Period and Model Precision}
\label{sec:Appendix_C}

In this study, we employ the computationally-demanding gradient descent method to estimate the model parameters.
Thus, to make the IP-CSM a practical model, we should not estimate the parameters at every step but at some finite interval.
In this section, we discuss the practical parameter estimation interval by evaluating the trade-off between the interval $\Delta t_{\mathrm{est}}$ and the estimated parameter values.

\begin{figure}[tb]
    \centering
    \begin{minipage}{0.45\textwidth}
        \centering
        \includegraphics[width=\textwidth]{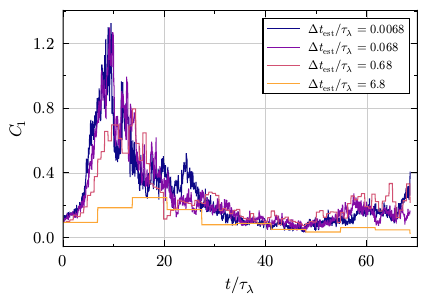}
        \subcaption{}
    \end{minipage}
    \begin{minipage}{0.45\textwidth}
        \centering
        \includegraphics[width=\textwidth]{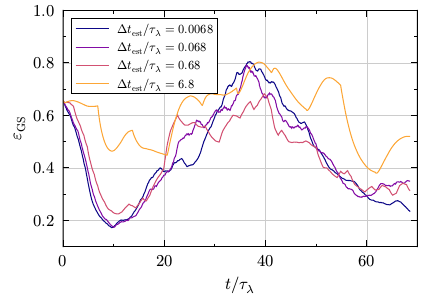}
        \subcaption{}
    \end{minipage}
    \caption{
        Time series of (a) the parameter $C_1(t)$ and (b) the grid-scale energy dissipation $\varepsilon_{\mathrm{GS}}(t)$ for different parameter estimation intervals $\Delta t_{\mathrm{est}}$.
        Time is normalized by the characteristic time scale $\tau_\lambda \coloneq u_{\mathrm{rms}} / \lambda = \sqrt{15\nu / \varepsilon}$ corresponding to the Taylor microscale $\lambda$.
        The interval is varied for 1, 10, 100, and 1000 time steps, each corresponding to $\Delta t_{\mathrm{est}} / \tau_\lambda = 0.0068$, 0.068, 0.68, and 6.8, respectively.
    }
    \label{fig:estimate_time_series}
\end{figure}

Figure~\ref{fig:estimate_time_series} shows the time series of the parameter $C_1(t)$ and the grid-scale energy dissipation $\varepsilon_{\mathrm{GS}}(t)$ for different parameter estimation intervals.
When the estimation interval satisfies $\Delta t_{\mathrm{est}} < \tau_\lambda$, where $\tau_\lambda \coloneq u_{\mathrm{rms}} / \lambda = \sqrt{15\nu / \varepsilon}$ is the Taylor microscale-based time scale, both the parameter and the statistical quantity fluctuations are generally collapsed.
On the other hand, when $\Delta t_{\mathrm{est}} > \tau_\lambda$, they exhibit smaller fluctuations and deviate from the results obtained with $\Delta t_{\mathrm{est}} < \tau_\lambda$.
These results suggest that the appropriate parameter estimation interval should be $\Delta t_{\mathrm{est}} \approx \tau_\lambda$.
In the main text, we conducted the IP-CSM simulation with the estimation interval $\Delta t_{\mathrm{est}} / \tau_\lambda = 0.68$, which corresponds to 100 time steps.

\end{document}